# SECURING INTERNET PROTOCOL (IP) STORAGE: A CASE STUDY

SIVA RAMA KRISHNAN SOMAYAJI[1] AND Ch.A.S MURTY[2]

[1]School of Information Technology and Engineering, VIT University, Vellore

somayaji1984@gmail.com

[2]Network and Security, C-DAC, Hyderabad

chasmurty@cdac.in

*ABSTRACT*

*Storage networking technology has enjoyed strong growth in recent years, but security concerns and threats facing networked data have grown equally fast. Today, there are many potential threats that are targeted at storage networks, including data modification, destruction and theft, DoS attacks, malware, hardware theft and unauthorized access, among others. In order for a Storage Area Network (SAN) to be secure, each of these threats must be individually addressed. In this paper, we present a comparative study by implementing different security methods in IP Storage network.*

## 1. INTRODUCTION

The proliferation of higher performing networks with multi-Gigabit Ethernet backbones, easier access to high-performance global networks such as Multiprotocol Label Switching (MPLS), and increasing popularity of Internet Simple Computer System Interface (iSCSI), an IP-based protocol which enables block-level I/O, IP storage networks are in dire need of secure transport which will not impact performance. In addition to storage performance, a practical IP- based security solution must also be simple, compatible, non-intrusive and cost- effective.

In a heterogeneous environment, we have the option of securing communication at both the application layer, using protocols such as Secure Sockets Layer (SSL) or the Transport Layer Security (TLS), and on the IP level using IPSec. The starting point for a systematic approach to storage security is to take stock of the various types of data being stored and classifying it according to how important it is and how costly it would be to the business if it were lost or stolen. Then for each classification, appropriate security policies should be set. The next step is to enforce password and World Wide name identification (for Fibre Channel) and logical unit number (LUN) authorization to ensure that only authorized users, devices or applications can access data, and to implement LUN masking so that particular storage volumes can only be seen by authorized users, devices or applications. iSCSI protocol and its related iSCSI drivers provide authentication features for both the initiator and target nodes. This can prevent unauthorized access and allow only trustworthy nodes to complete communications.

In order to transfer data to and from the storage securely on an iSCSI network, iSCSI can employ Ipsec that offers strong encryption and authentication functions for IP packets. However, the encryption processing triggers performance degradation when mass volume of data should be transferred. Specifically in a long-latency environment, ACK or a SCSI Command takes a long time until it arrives at the other machine. Moreover, Ipsec is implemented in IP layer located on the lower-level. If we try to improve the performance of Ipsec encryption processing, IP and other codes inside a kernel of operating systems are required to be modified.

In this paper, we have shown the performance analysis of IP storage network in different scenarios.

## 2. RELATED WORK

There has been lot of work done in the implementation of IP-storage. Soumen Debgupta [1] proposes a software approach of iSCSI by exploiting the optional features like multiple connections to improve performance. Yi-Cheng [2] presented a methods for implementing the implementation of the iSCSI virtualization switch used in SANs. The proposed method reduces the overheads of protocol processing by using a packet forwarding model based on caching the structure ID of the iSCSI session. Dimitar[3] proposed that iSCSI host bus adapters, also called iSCSI NICs or Storage NICs (SNIC), are optimized in hardware with realization of a TCP/IP off-load engine (TOE) to minimize processing overhead to achieve better performance in IP-SAN. Kamisaka [4] presented a method of optimization for encryption processing in the upper-layer instead of using Ipsec.

Dr. Rekha Singhal [5] proposes two novel techniques for improving the performance of iSCSI protocol. First proposed technique is the elimination technique for reducing latency caused by redundant overwrites and the second technique reduces the latency caused due to multiple reads to the same storage sector. Dr. Zia Saquib[6] propose a method of using clusters of inexpensive nodes with Redundant Array of Inexpensive Nodes using iSCSI for high performance using commodity hardware and setting up efficient iSCSI target controllers for block virtualization.

## 3. iSCSI PROTOCOL MODEL

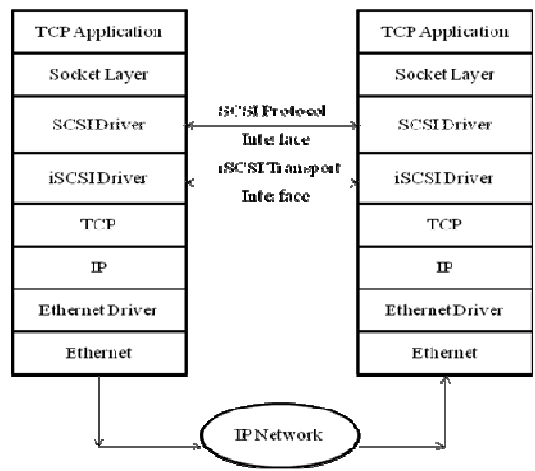

Figure 1: IP storage layered model

The above figure shows how a communication is taken place by an initiator and target. The iSCSI system is a layered structure consisting of SCSI/iSCSI and TCP/IP.

International Journal of Next Generation Network (IJNGN), Vol.2, No.1, March 2010

### 3.1. Details of the Initiator

In the implementation we have used Windows Vista systems as the initiators and target. In Windows Vista, the iSCSI initiator driver software is readily available.

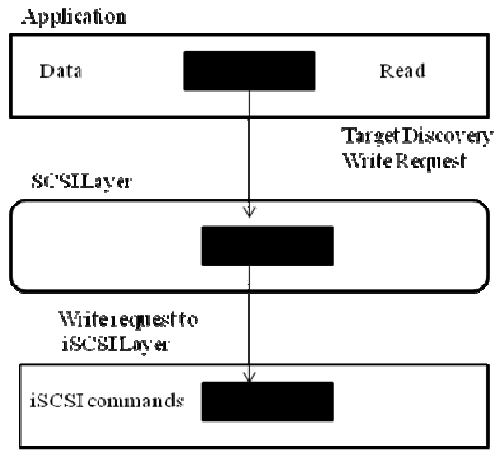

Figure 2: Data processing in Initiator

Figure 2 describes the basic model of how data is processed in the initiator. First the initiator searches for targets available. This is the discovery phase. When the initiator discovers a target, a data write request command is initialized and data is sent to lower iSCSI/SCSI layer where iSCSI commands are processed and then the data is sent to the appropriate target.

### 3.2. DETAILS OF TARGET

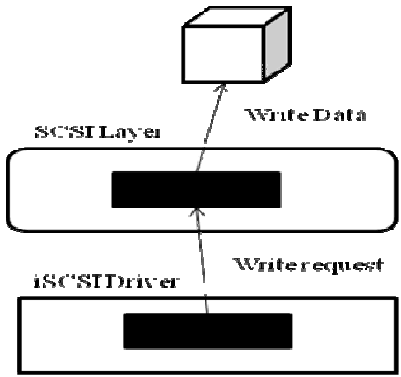

Figure 3: Data processing in Target

Figure 3 describes how the data is sent to the disk arrays. The data comes from the TCP layer to SCSI/iSCSI layer where a write request is called. The data segments are passed to the *handle cmd* function at iSCSI/SCSI driver and they are written in the target's disk sequentially.

## 4. PERFORMANCE ANALYSIS OF IP-STORAGE NETWORK

### 4.1 Without any Security Implementation

The traffic analysis is done using a tool *wireshark* which is a open source and a free downloadable software for protocol analysis.

Figure 4: Traffic analysis between initiator and the target

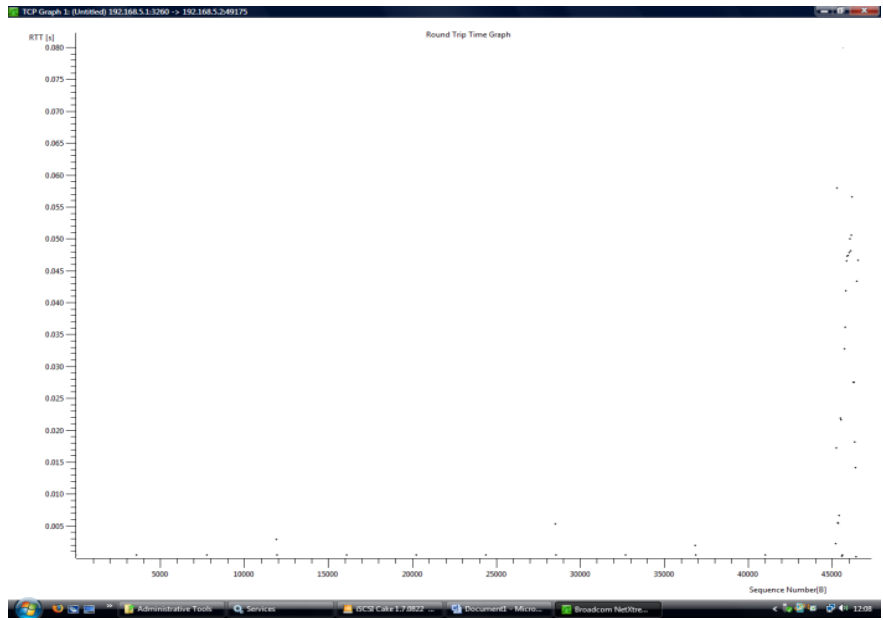

Figure 5: Round Trip Time Graph

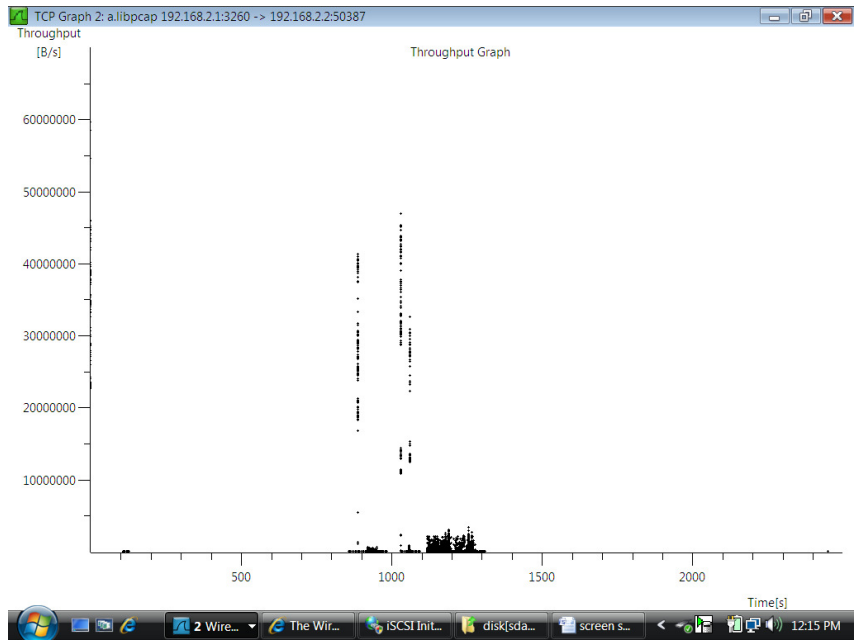

Figure 6: Throughput Graph

## 4.2. SSLv2 Implementation.

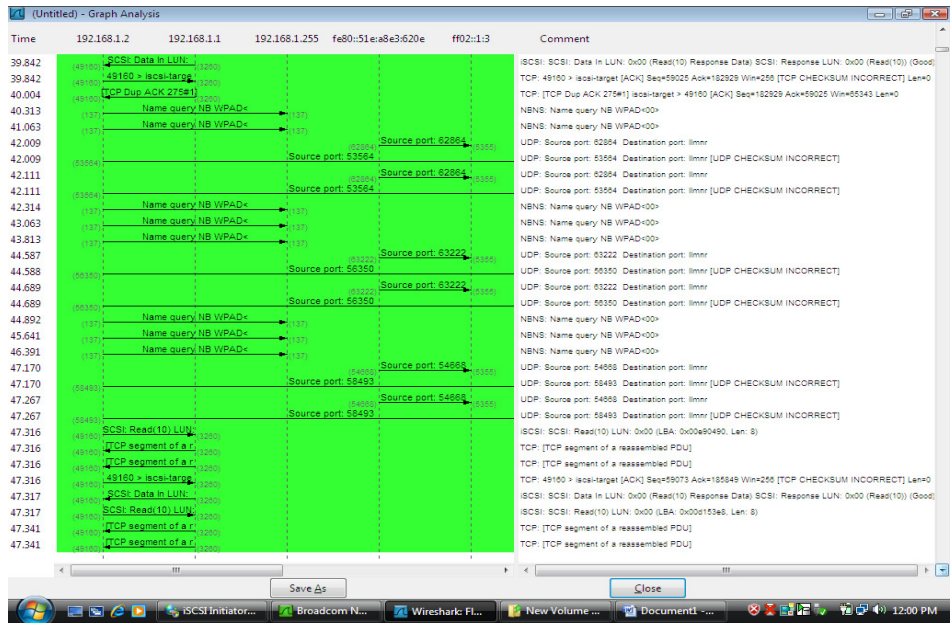

Figure 7: Graph Analysis with SSLv2 enabled in IP-Storage

In the above figure, the initiator contacts its local system Name query through port 137 and at the source port 53564 the encryption process is started at the initiator and at port 62864 the UDP checksum is performed by link local multicast name resolution at the destination.

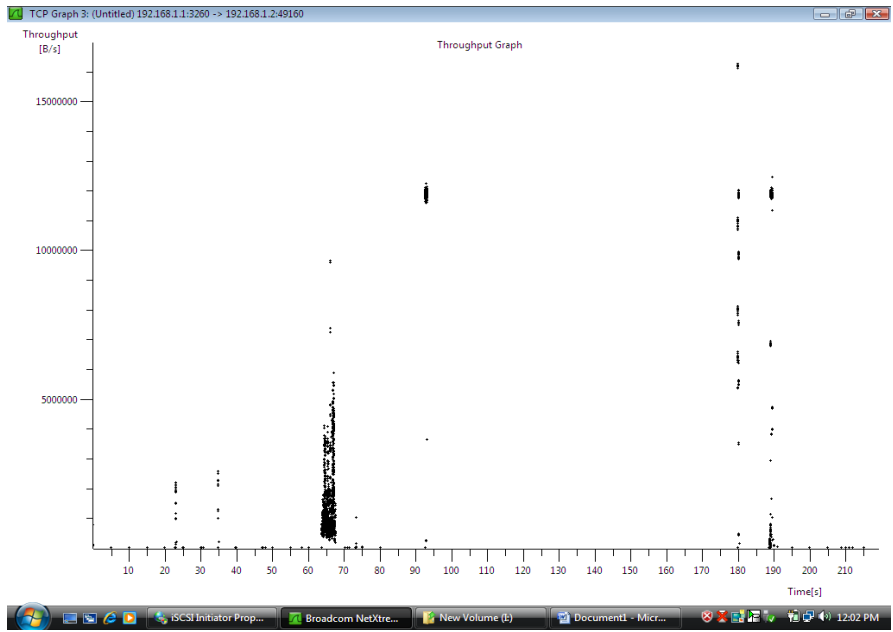

Figure 8: Throughput Graph

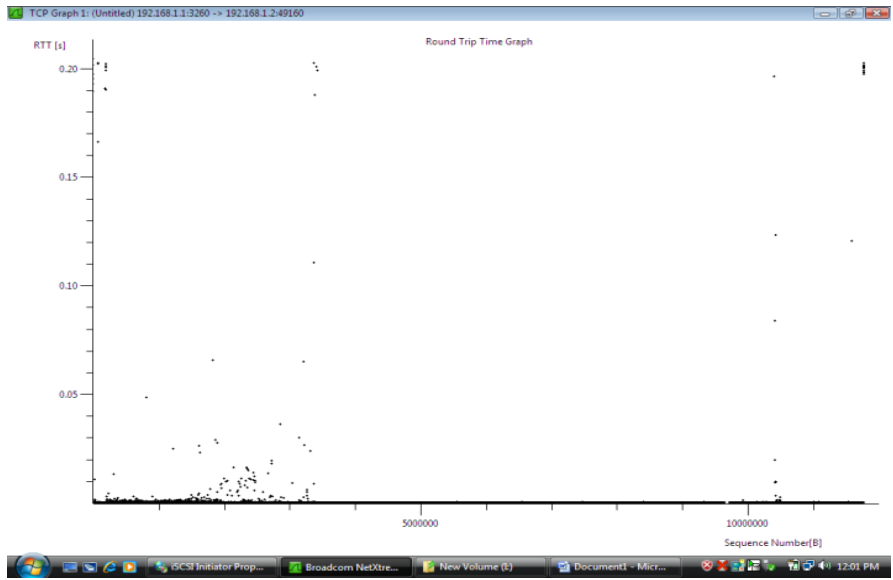

Figure 9: Round trip time graph

When we compare figure 8 (Throughput graph) and figure 9 (Round trip time graph) we notice that there is a slight decrease in the performance in the IP-storage network. This decrease is due to the fact that SSL generates some source code for every application.

### 4.3 IP-sec Implementation.

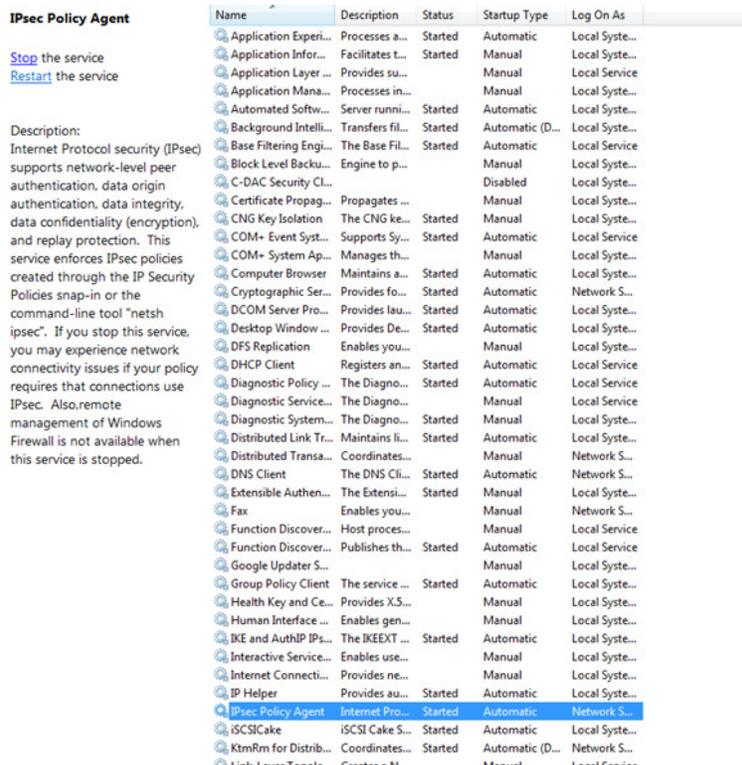

Figure 10: Enabling IPsec Policy Agent

IPsec can be enabled by msc services. We can find the IPsec policy disabled. Starting this service enables IPsec.

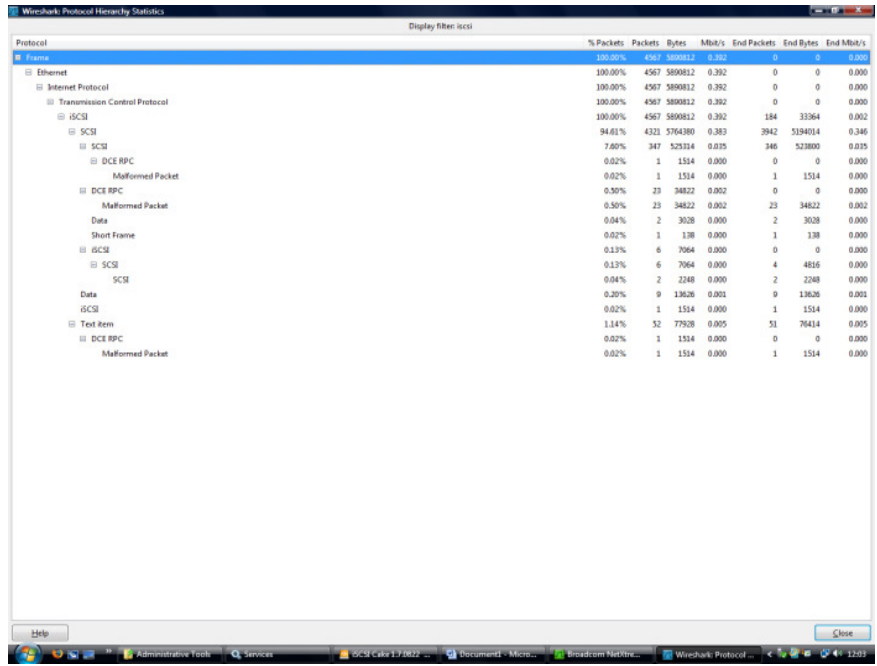

Figure 11: Protocol Hierarchy Statistics

In the above figure. A remote procedure call has been invoked called the DCE/RPC (Distributed Computing Environment / Remote Procedure Calls).

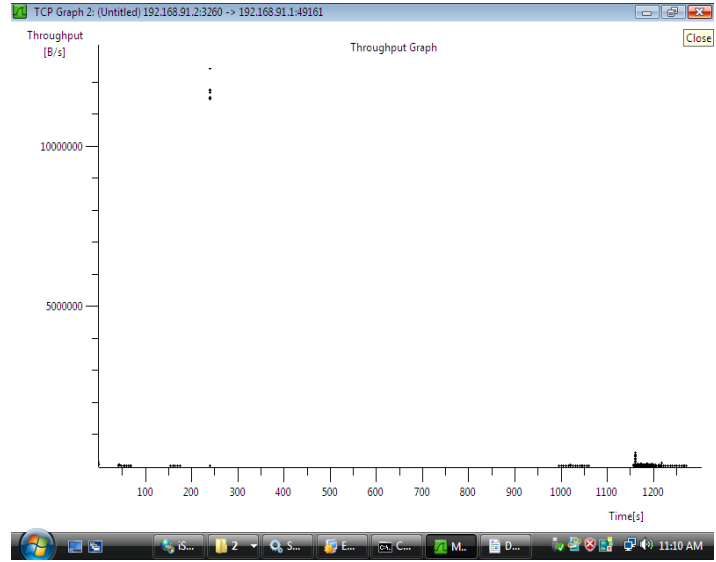

Figure 12: Throughput graph

In the above throughput graph, there is drastic performance degradation after 1000 seconds.

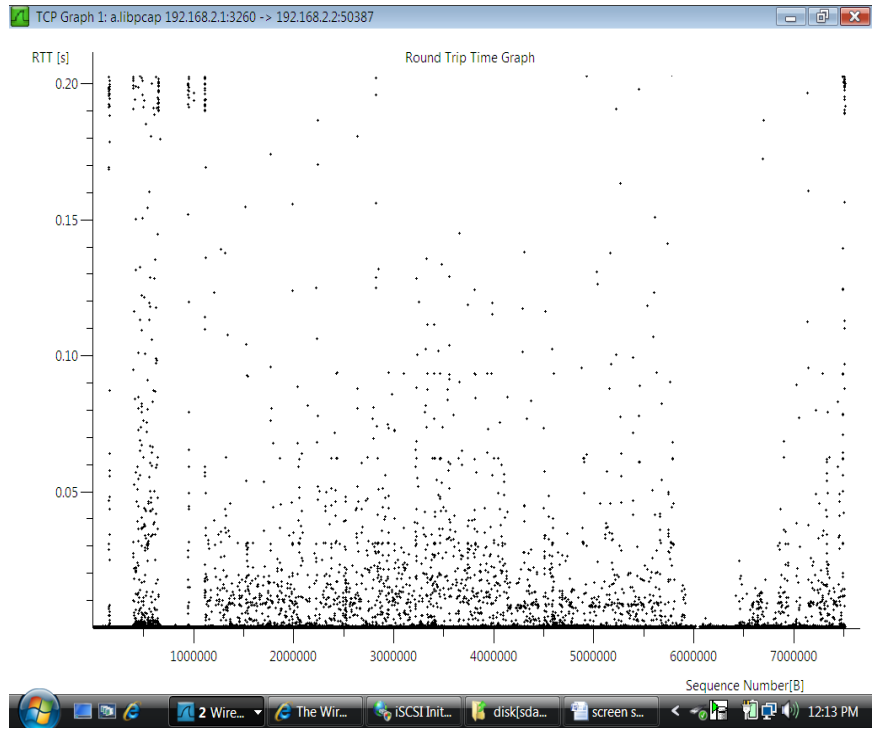

Figure 13: Round trip time graph

The above figure clearly tells us that that is much delay when we implement IP-sec as there are lots of peaks (bottle-necks) at different time instances as compared to the SSLv2 implementation.

## 5. COMPARISON OF PERFORMANCE ANALYSIS OF IMPLEMENTATION OF SSLv2 AND IPSEC

|  | Round trip time values (Sequence no, Time) | Throughput values (Time, No of bits) |
|---|---|---|
| SSLv2 | (1000, 0.01) | (25, 1000) |
|  | (3000, 0.005) | (70, 5000) |
|  | (5000, 0.07) | (90, 12000) |
|  | (10000, 0.02) | (190, 20000) |
|  | Round trip time values | Throughput values |
| IPsec | (1000, 0.01) | (20, 1000) |
|  | (3000, 0.03) | (70, 50000) |
|  | (5000, 0.057) | (90, 100000) |
|  | (10000, 0.12) | (190, 300000) |

Table: Comparative values of Round trip time graph and throughput graph

The above table shows that when we implement SSLv2 there is a decrease in the Round Trip Time and an increase in the throughput as compared to implementation of IPsec in the storage network. This is due to the fact that IPsec is implemented in the lower layers along with IP protocol and the IP needs to perform addition function i.e. securing the packets and then route them. In case of Secure Socket Layer (SSLv2), the security is implemented in sockets or at the port level and is transparent to the end application.

## 6. CONCLUSION

In this paper we have implemented an IP-Storage network using iSCSI protocol. We have analyzed the performance of the IP Storage network without any security implemented and also by implementing SSLv2 and IPsec. We present a comparative analysis IP storage network performance in each case.

## ACKNOWLEDGMENT

The authors thank the Dr.Sarat Chandra Babu, Director, C-DAC, for his encouragement and giving permission to publish this paper. We also thank to Prof. H.R.Vishwakarma, School of Computing Sciences, VIT University, Vellore for his continuous support and guidance.